\documentclass[aps,prl,10pt,twocolumn]{revtex4-1}

\usepackage{amsmath}
\usepackage{mathtools}
\usepackage{graphicx}
\usepackage{amssymb}
\usepackage{times}
\usepackage{bbold}
\usepackage{color}
\usepackage{braket}

\usepackage{hyperref}

\hypersetup{
colorlinks=true,
linkcolor=blue,          
citecolor=blue,          
filecolor=magenta,       
urlcolor=cyan
}

\DeclareMathOperator{\Tr}{Tr}

\begin{document}

\title{Robustness of Majorana edge modes and topological order --- \\exact results for the symmetric interacting Kitaev chain with disorder}

\author{Max McGinley}
\affiliation{Cavendish Laboratory, University of Cambridge, Cambridge CB3 0HE, United Kingdom}
\author{Johannes Knolle}
\affiliation{Cavendish Laboratory, University of Cambridge, Cambridge CB3 0HE, United Kingdom}
\author{Andreas Nunnenkamp}
\affiliation{Cavendish Laboratory, University of Cambridge, Cambridge CB3 0HE, United Kingdom}

\date{\today}

\begin{abstract}
We investigate the robustness of Majorana edge modes under disorder and interactions. We exploit a recently found mapping of the {\it interacting} Kitaev chain in the symmetric region ($\mu = 0$, $t = \Delta$) to free fermions. Extending the exact solution to the disordered case allows us to calculate analytically the topological phase boundary for all interaction and disorder strengths, which has been thought to be only accessible numerically. We discover a regime in which moderate disorder in the interaction matrix elements enhances topological order well into the strongly interacting regime $U > t$. We also derive the explicit form of the many-body Majorana edge wave function revealing how it is dressed by many-particle fluctuations from interactions. The qualitative features of our analytical results are valid beyond the fine-tuned integrable point as expected from the robustness of topological order and as corroborated here by an exact diagonalization study of small systems.
\end{abstract}

\maketitle

\bibliographystyle{apsrev4-1.bst}

Majorana edge modes in condensed matter physics have recently received a great deal of attention \cite{Alicea2012} primarily due to their applications in topological quantum computation \cite{Nayak2008}. In a seminal paper \cite{Kitaev2001} Kitaev introduced the minimal model of a 1D $p$-wave superconducting wire, now known as the Kitaev chain. One of its remarkable properties is the presence of zero-energy states localized at the two ends of the chain. Paired together, these Majorana edge modes can form a qubit which is largely protected from decoherence due to its non-local nature.

Compelling experimental evidence of their existence has been reported in semiconducting nanowires in proximity to $s$-wave superconductors \cite{Mourik2012, Deng2012, Das2012, Deng2016} and in ferromagnetic atomic chains \cite{NadjPerge2014, Ruby2015, Pawlak:2016}. However, there remains a possibility that the zero-bias conductance peak measured in these experiments is due to disorder rather than due to Majorana modes \cite{Liu2012, Rainis:2013, Sau2013}, and so it is important to include disorder in theoretical investigations. Additionally, the nature of these experimental platforms inevitably leads to the presence of interactions between the low-energy degrees of freedom \cite{Stoudenmire2011, Sela2011, Else2017}.

The majority of analytical studies of the Kitaev chain have focussed on the clean, non-interacting case \cite{Alicea2012}. Beyond this, for clean, interacting chains, only few exact results are known \cite{Gangadharaiah:2011, Altland2014, Katsura2015, Fendley2016}, and numerical/perturbative studies have shown that Majorana edge modes can be stable up to moderate interaction strengths \cite{Stoudenmire2011, Sela2011, Hassler:2012, Thomale2013, Chiu2015}. Similarly, a number of works on non-interacting, disordered/quasi-periodic chains find a relatively broad parameter region of stability \cite{Huse2001, Degottardi2013, Cai2013}. The combined effect of interactions and disorder in Kitaev chains has recently been studied numerically \cite{Crepin:2014, Gergs2016}, as well as through a weak-disorder renormalisation group approach \cite{Lobos2012}. However, an analytic treatment of both strong interactions and strong disorder has been thought to be impossible.

\begin{figure}
\includegraphics[scale=0.9]{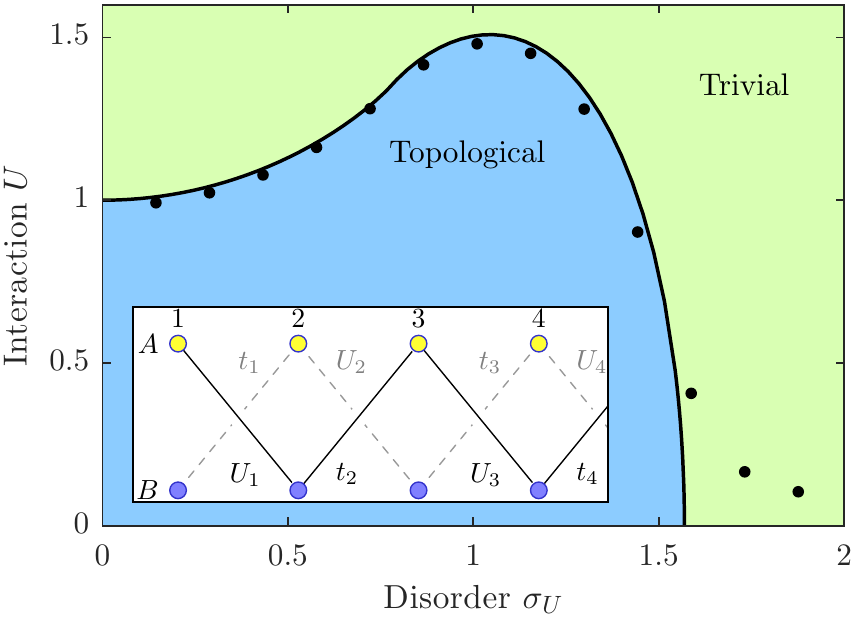}
\caption{Topological phase diagram of Hamiltonian (\ref{eqHamiltOrigGeneral}) as a function of the mean interaction $U$ and disorder strength $\sigma_U$. Solid line is the exact boundary given by the analytic condition Eq.~\eqref{eqPhaseBoundary}. Points show the critical interaction $U$ using the entanglement degeneracy criterion (see text) calculated numerically for $N = 800$ sites, averaged over 500 disorder realizations. Error bars are smaller than the symbol size. Inset: Visual representation of the decoupling of Hamiltonian \eqref{eqHamiltTransformed} into two subsystems. Solid (dashed) lines depict terms in subsystem I (II).}
\label{figTopTrans}
\end{figure}

In this paper we investigate analytically the combined effects of disorder and interactions on topological order. Focussing on the example of a Kitaev chain supplemented with disorder and density-density interactions, we exploit a recently found exact solution of Miao {\it et al.}~\cite{Miao2016} to classify its topological order and to construct the Majorana edge modes explicitly. The solution is valid in the symmetric region, which is particle-hole symmetric $\mu = 0$ and has hopping amplitude and superconducting pairing equal on each site $t_j = \Delta_j$. Exact results are obtainable for any configuration of hopping/pairing amplitudes and interaction strengths, allowing us to access the strongly interacting and disordered regions of the phase diagram.

By considering the normalizability of the topological edge modes, we are able to obtain an analytic condition for the ordered phase for arbitrary disorder distributions. Focussing on the specific case of uniform disorder, we calculate the phase diagram as a function of the mean and width of the disorder (Fig.\@ \ref{figTopTrans}).
We find that moderate disorder enhances the topological phase well into the strongly disordered and interacting regime $U \gtrsim t$. This suppression of the interaction-driven destruction of Majorana edge states has been inaccessible to previous numerical and RG treatments. 

\begin{figure}
\includegraphics[scale=0.9]{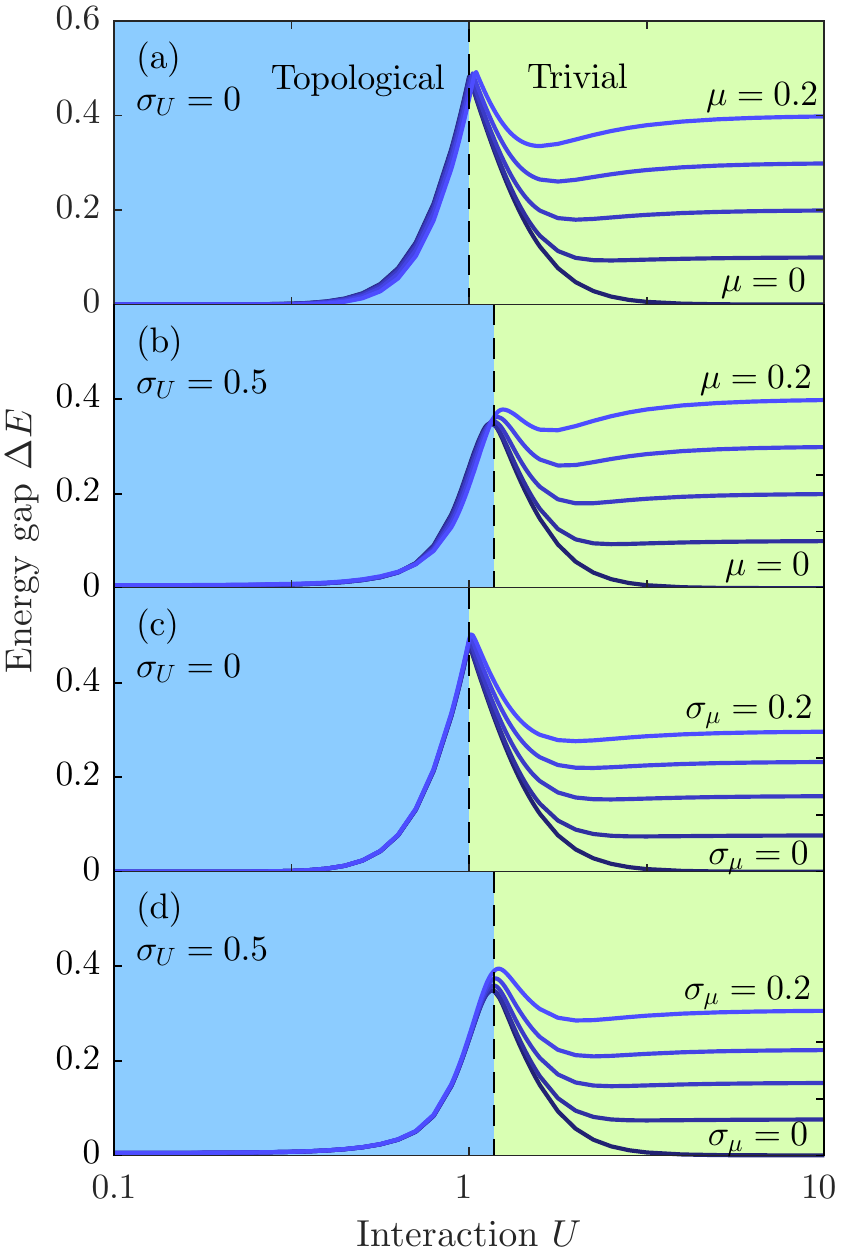}
\caption{Energy gap $\Delta E$ of the Hamiltonian (\ref{eqHamiltOrigGeneral}) by ED of $N = 12$ sites as a function of mean interaction $U$ with $t_j=\Delta_j=1$. Uniform chemical potential $\mu_j = \mu$ from $\mu=0$ to $\mu=0.2$ for (a) clean $\sigma_U=0$ and (b) disordered $\sigma_U=0.5$ case. Disordered chemical potential $\mu_j \in [-\sqrt{3} \sigma_\mu, \sqrt{3}\sigma_\mu]$ from $\sigma_\mu=0$ to $\sigma_\mu=0.2$ for (c) $\sigma_U=0$ and (d) $\sigma_U=0.5$, respectively. In (b), (c), and (d), we have averaged over 500 disorder realizations.}
\label{figMajInstab}
\end{figure}

Whilst our exact results are obtainable only in the symmetric region, we show by exact diagonalization (ED) that the qualitative aspects of our findings hold away from this fine-tuned limit (Fig.\@ \ref{figMajInstab}), indicating that the predicted phenomena could be observed in current nanowire-based experiments. Moreover, experiments on quantum-dot chains \cite{Sau2012} can be deliberately tuned to the region of interest $t_j = \Delta_j$, $\mu = 0$ \cite{Fulga2013}.

\emph{Model and Majorana modes.---}
We consider a generalization of the Kitaev chain describing spinless fermions on a one-dimensional lattice with open boundary conditions. The Hamiltonian is
\begin{align}
H &= \sum_{j=1}^{N-1} \left[-t_j(c^\dagger_j c_{j+1} + \text{h.c.}) -\Delta_j(c^\dagger_j c^\dagger_{j+1} + \text{h.c.}) \right.\nonumber\\&- \left.\mu_jc_j^\dagger c_j + U_j(2c_j^\dagger c_j - 1)(2c_{j+1}^\dagger c_{j+1}-1)\right],
\label{eqHamiltOrigGeneral}
\end{align}
with a hopping term $t_j$, an on-site chemical potential $\mu_j$, a $p$-wave superconductor pairing $\Delta_j$, and a nearest-neighbor interaction $U_j$. Kitaev's original model had $U_j = 0$, so we refer to $U_j \neq 0$ as an \textit{interacting} Kitaev chain.

The topological properties of our system are best studied using a basis of Majorana operators, defined as $\gamma_j^A = c_j + c_j^\dagger$ and  $\gamma_j^B = -i(c_j - c_j^\dagger)$ which satisfy the anti-commutation relation $\{\gamma_j^\alpha, \gamma_k^\beta\} = 2\delta_{j,k} \delta^{\alpha, \beta}$. The clean non-interacting model has two gapped phases which differ by the existence of a pair of Majorana modes localized at either edge. Such a mode can be described by a Hermitian operator $Q$ which commutes with the Hamiltonian $[H, Q] = 0$ (up to corrections that decay exponentially with the system size). As an example, if $\mu_j = U_j = t_j-\Delta_j=0$, we get $Q = \gamma_1^A$ or $\gamma_N^B$, each of which are absent from the Hamiltonian and so commute exactly. Within this phase, the action of either $Q$ operator on any eigenstate produces a different eigenstate of the same energy, and so the energy spectrum is doubly degenerate. Importantly, the two states in each pair have opposite fermion number parity, i.e.~are eigenstates of the operator
\begin{align}
Z_2^c = (-1)^{\sum_{j=1}^N c_j^\dagger c_j}
\label{eqZ2c}
\end{align}
(which itself commutes with $H$), with eigenvalues $\pm 1$, corresponding to even and odd numbers of fermions, respectively. This non-local observable acts as a topological order parameter distinguishing between even and odd sectors. If degenerate states have opposite $Z_2^c$, then $Q$ must anticommute with $Z_2^c$. In this case, the phase is topologically ordered, with twofold degenerate ground states in opposite topological sectors.

Topological order is robust against local perturbations, and so when we include interactions in our system, we expect the Majorana edge modes to persist up to some critical interaction strength. Whilst the microscopic nature of the Majorana modes will inevitably be different from the non-interacting case, they should still satisfy the same requirements of being localized at either edge, commuting with the Hamiltonian, and anticommuting with $Z_2^c$ \cite{Fendley2012}. However, unlike the non-interacting case, if $U_j \neq 0$ then $Q$ cannot be written as a linear combination of $\gamma$ operators because the Majorana modes will be dressed by higher-order multiple-particle contributions. In general, the Majorana mode generalizes to a many-body Majorana operator with an expansion \cite{Kells2015}
\begin{align}
Q = \sum_{\substack{j\\ c = A,B}} \alpha^j_c \gamma_j^c + \sum_{\substack{j, j', j''\\ c, c', c'' = A,B}} \alpha^{j,j',j''}_{c, c', c''} \gamma_j^c \gamma_{j'}^{c'} \gamma_{j''}^{c''} + \cdots
\label{eqMajGeneral}
\end{align}
featuring terms with an odd number of Majorana operators. If all the coefficients $\alpha$ are real, then this describes a Hermitian operator that anticommutes with $Z_2^c$. In the topological phase, we can find two normalizable $Q$ operators which commute with $H$, one localized at each edge (in the sense that $\alpha_{\ldots,j,\ldots} \rightarrow 0$ as $j \rightarrow \infty$ for the left mode, and similarly for the right mode).

Constructing explicit expressions for the $\{\alpha\}$ is generally only possible numerically due to the complexity of the many-body problem. However, here for the symmetric chain we derive for the first time closed expressions for the coefficients of the many-body Majorana operator \eqref{eqMajGeneral} for an interacting disordered system. This allows us to classify the phases as topological if the many-body Majorana operator exists.

\emph{Exact solution.---}
To achieve this, we make use of an exact solution due to Miao \textit{et al.\@} \citep{Miao2016}. Using two successive Jordan-Wigner transformations and a spin rotation, they showed that for the clean case in the symmetric region $\mu = 0$ and $t = \Delta$, the non-local transformation
\begin{align}
\lambda_j^A &=
\begin{dcases*}
\left( \prod\nolimits_{k\, \text{odd}}^{j-1} i\gamma_{k}^B \gamma_{k+1}^A\right)\gamma_{j}^A & $j$ odd;\\
 \left(\prod\nolimits_{k\, \text{odd}}^{j-3} i\gamma_{k}^A \gamma_{k+1}^B\right)(i\gamma_{j-1}^A \gamma_{j}^A) & $j$ even;
\end{dcases*} \nonumber\displaybreak[0]\\
\lambda_j^B &=
\begin{dcases*}
\left(\prod\nolimits_{k\, \text{odd}}^{j-2} i\gamma_{k}^A \gamma_{k+1}^B\right)(i\gamma_{j}^A \gamma_{j}^B) & $j$ odd;\\
\left(\prod\nolimits_{k\, \text{odd}}^{j-1} i\gamma_{k}^B \gamma_{k+1}^A\right)\gamma_{j}^B & $j$ even
\end{dcases*}
\label{eqFullTrans}
\end{align}
preserves the Majorana anticommutation relations, so we have $\{\lambda_j^\alpha, \lambda_k^\beta\} = 2\delta_{j,k}\delta^{\alpha, \beta}$. This allows us to express the Hamiltonian (\ref{eqHamiltOrigGeneral}) in terms of $\lambda$-fermion bilinears. We note that this also holds for disordered $t_j$ and $U_j$, yielding the Hamiltonian
\begin{align}
H = \sum_{j=1}^{N-1}\left[-it_j\lambda_{j+1}^A \lambda_j^B + iU_j \lambda_j^A\lambda_{j+1}^B\right].
\label{eqHamiltTransformed}
\end{align}
We depict the Hamiltonian (\ref{eqHamiltTransformed}) visually in the inset of Fig.\@ \hyperref[figTopTrans]{\ref{figTopTrans}}, using lines to represent fermion bilinears. It is evident that one half of the Majorana operators decouples from the other half, so we can consider two subsystems separately, which we label with Roman numerals I and II. We make this explicit by redefining
$\phi_{\text{I},j}^A = \lambda_{2j-1}^A,\; \phi_{\text{I},j}^B = \lambda_{2j}^B ,\;
\phi_{\text{II},j}^A = \lambda_{2j-1}^B ,\; \phi_{\text{II},j}^B = \lambda_{2j}^A$,
so the Hamiltonian is the sum of two uncoupled chains
\begin{align}
H &= \sum_{j=1}^{N/2} \left[ -i t_{2j} \phi_{\text{I},j+1}^A \phi_{\text{I},j}^B + iU_{2j-1} \phi_{\text{I},j}^A \phi_{\text{I},j}^B \right] \nonumber\\&+ \sum_{j=1}^{N/2} \left[ i U_{2j} \phi_{\text{II},j+1}^A \phi_{\text{II},j}^B - it_{2j-1} \phi_{\text{II},j}^A \phi_{\text{II},j}^B \right].
\label{eqTransTF}
\end{align}
Each of the two subsystems is equivalent to a non-interacting Kitaev chain of length $N/2$ with $t_j = \Delta_j$, one of which has the parameters $\mu_j \rightarrow 2U_{2j-1}$ and $t_j \rightarrow t_{2j}$, and the other of which has the parameters $\mu_j \rightarrow -2t_{2j}$ and $t_j \rightarrow -U_{2j-1}$. \textit{Mutatis mutandis}, from our knowledge of non-interacting Kitaev chains, we can identify quantum phase transitions in the clean case at $U = \pm t$ at which one subsystem becomes topological and the other becomes trivial in this transformed basis. We note that this decoupling is analogous to the equivalence between the $XY$ model and two independent transverse-field Ising models \cite{Jullien1978}, as each can be related to our system via Jordan-Wigner transformations. 

\emph{Topological phase boundary.---} Having reduced the original interacting Hamiltonian (\ref{eqHamiltOrigGeneral}) to a quadratic one (\ref{eqTransTF}), we know that any zero-energy boundary mode in the new basis is a linear combination of single fermion operators
\begin{align}
Q^A_\text{I} = \sum_j \alpha^j_\text{I} \phi_{\text{I},j}^A && Q^A_\text{II} = \sum_j \alpha^j_\text{II} \phi_{\text{II},j}^A
\end{align}
with similar expressions for $Q^B_\text{I}$ and $Q^B_{\text{II}}$. Because the system after the non-linear transformation is equivalent to a conventional non-interacting Kitaev chain, we can use the standard expression for a non-interacting boundary mode with $t = \Delta$ \cite{Fendley2012}, with the appropriate reassignments of $\mu$ and $t$, giving us $\alpha_\text{I}^j \propto (-U / t)^{j-1}$ and $\alpha_\text{II}^j \propto (-t/U)^{j-1}$. Within each phase, only one of the subsystems has a normalizable mode $Q^2 < \infty$.

We can transform these operators back into the original basis using Eq.~(\ref{eqFullTrans}). In the $|U| < t$ phase, subsystem I possesses Majorana  modes and the {\it many-body Majorana operator} is
\begin{align}
Q^A_\text{I} &= \alpha^1_\text{I} \gamma_1^A + \alpha^2_\text{I} \gamma_1^B (i\gamma_2^A \gamma_3^A) \nonumber\\ &+ \alpha^3_\text{I} \gamma_1^B (i\gamma_2^A \gamma_3^B) (i\gamma_4^A \gamma_5^A) + \cdots.
\label{eqMajInteracting}
\end{align}
This expression is a generalization of a non-interacting Majorana mode and a special case of Eq.~\eqref{eqMajGeneral} for which the coefficients $\alpha$ can be given explicitly. It is an edge mode in the sense that terms featuring the operators $\gamma_j^{A,B}$ decay exponentially with $j$, and again it is a Majorana operator since it is Hermitian and anticommutes with the fermion parity operator $Z_2^c$. Additionally, the mode is adiabatically connected to a non-interacting Majorana mode -- for $U \rightarrow 0$ all multi-particle terms vanish leaving us with a single $\gamma$ operator.

On the other hand, when $|U| > t$, the edge mode changes to $Q^A_\text{II}$ which in the original basis is
\begin{align}
Q^A_\text{II} &= \alpha^1_\text{II} (i\gamma_1^A \gamma_1^B) + \alpha^2_\text{II} (i\gamma_1^A \gamma_2^B)(i\gamma_3^A \gamma_3^B) \nonumber\\ & +\alpha^3_\text{II} (i\gamma_1^A \gamma_2^B)(i\gamma_3^A \gamma_4^B) (i\gamma_5^A \gamma_5^B) + \cdots.
\label{eqMajInteractionTrivial}
\end{align}
It has a form similar to that of the $Q^A_\text{I}$ mode (\ref{eqMajInteracting}), but with the crucial difference that it \textit{commutes} with $Z_2^c$ and does not have the form of Eq.~\eqref{eqMajGeneral}. Therefore, Equation (\ref{eqMajInteractionTrivial}) cannot represent a topological edge mode, and cannot be adiabatically connected to any other Majorana mode. Acting on states with $Q_\text{II}^A$ does indeed generate different states of the same energy, but this is an accidental degeneracy of the symmetric chain (specifically $\mu = 0$), so arbitrarily small perturbations from the fine-tuned point will destroy the degeneracy and the edge mode, as shown below by ED. We thus classify this phase as topologically trivial.

\emph{Disordered phase diagram.---}
We now consider the case where the parameters $U_j$ and $t_j$ are sampled from probability distributions $P(U)$ and $P(t)$. To calculate the wavefunction coefficients $\alpha_\text{I}^j$ and $\alpha_\text{II}^j$, we impose the condition $[H, Q] = 0$, as done for the non-interacting disordered case \cite{DeGottardi2011}, yielding $
\alpha^{j+1}_\text{I} = -(U_{2j-1}/t_{2j})\alpha^{j}_\text{I}$.
For such a mode to exist we will need to be able to normalize it, i.e.~$(Q^A_\text{I})^2 = 1$, and thus the sums of the squares of the coefficients $\alpha^j$ need to be bounded. As before, the condition for the topological phase is that $Q^A_\text{I}$ exists, and so $\alpha^j_\text{I}$ must decay sufficiently fast to the right $j\rightarrow \infty$. Specifically,
\begin{align*}
\lim_{N \rightarrow \infty} \; \sum_{j=1}^N \left[ \prod_{k=1}^j \left|\frac{U_{2k-1}}{t_{2k}}\right|^2 \right] < \infty.
\end{align*}
Repeating the argument for $B$-flavor modes, we find that the above is also the condition for $Q^B_\text{I}$ to be localized on the right. The $j^\text{th}$ term $S_j$ in the sum above can be written as $S_j = \exp(\sum_{k=1}^j 2\ln|U_{2k - 1}/t_{2k}|)$ which tends to  $e^{2j\langle \ln U - \ln t \rangle}$ as $j \rightarrow \infty$, as the sum is self-averaging. Clearly, if
\begin{align}
\int dU P(U) \ln |U| < \int dt P(t) \ln |t|
\label{eqMajNorm}
\end{align}
then $S_j$ decays exponentially with $j$ and the sum converges. Equation \eqref{eqMajNorm} is the condition for the topological phase in the disordered system, which represents one of our main results. Within this phase, the Majorana modes have the interacting form \eqref{eqMajInteracting} with coefficients as calculated above. The argument above also gives us the characteristic decay length of the Majorana mode as $\xi = (\langle \ln t \rangle - \langle \ln U \rangle)^{-1}$, which for uniform or Gaussian distributions diverges as $\sim |\langle U \rangle - \langle U_\text{crit} \rangle |^{-1}$ at the phase transition.

Having derived the condition for the topological phase for arbitrary disorder distributions \eqref{eqMajNorm}, let us consider the specific example of constant $t_j = t$ and a uniform distribution for $U_j \in U + [-\sqrt{3} \sigma_U, \sqrt{3}\sigma_U]$, in units for which $t = 1$. We can construct the topological phase diagram as a function of $U$ and $\sigma_U$ by solving Eq.~\eqref{eqMajNorm} for a uniform $P(U)$, giving an analytic expression for the topological-trivial phase boundary
\begin{align}
&\left(\frac{U}{\sqrt{12}\sigma_U} + \frac{1}{2}\right)
\ln \left|U + \sqrt{3}\sigma_U\right| \nonumber\\
&- \left( \frac{U}{\sqrt{12}\sigma_U} - \frac{1}{2}\right)
\ln \left|U - \sqrt{3}\sigma_U\right| = 1,
\label{eqPhaseBoundary}
\end{align}
where we fixed $U > 0$ as the phase diagram is invariant under $U_j \rightarrow -U_j$. The critical $U$ grows quadratically for weak disorder and reaches a maximum at $(\sigma_U \approx 1.0451,\, U \approx 1.5089)$. We note that this is a significantly enhanced maximal interaction strength $U$ for which the phase is topological. For $\sigma_U > 2e/\sqrt{12} \approx 1.5694$ the system is trivial regardless of the mean interaction $U$. In this large disorder regime, the system is dominated by sites where $U_j$ is particularly large in magnitude, favoring a trivial charge ordered state.

\emph{Numerical results.---}
We corroborate our analytic results by calculating the phase diagram numerically. We compute the entanglement spectrum which is degenerate in a topological phase \cite{Turner2011}, see SI \cite{SI}. Specifically, we classify the phase as topological if the finite-size splitting of entanglement eigenvalues is less than some constant $c$. This criterion has been shown to be robust against disorder \cite{Gergs2016}.

Figure \ref{figTopTrans} shows the phase diagram of the disordered, interacting Kitaev chain, as given by the analytic expression \eqref{eqPhaseBoundary} as well as from the degeneracy of the entanglement spectrum for a system of size $N = 800$. We have set $c=0.1$ and verified that the transition is sharp enough to be insensitive to this arbitrary choice when averaged over $500$ disorder realizations. Finite-size effects lead to a slight suppression of the topological phase in the numerically calculated values, due to cases where the Majorana decay length is comparable to the system size. The transition to the trivial phase at strong disorder is also less sharp and finite-size errors are amplified.
In the SI \cite{SI} we demonstrate that these results for the symmetric region are robust to a disordered hopping amplitude.

Away from the symmetric region, we perform ED of the interacting, disordered Kitaev chain (\ref{eqHamiltOrigGeneral}). In Fig.~\ref{figMajInstab} we show the ground-state energy gap $\Delta E$ as a function of the mean interaction $U$ in the clean $\sigma_U = 0$ and disordered $\sigma_U \not= 0$ case, for uniform $\mu_j = \mu$ as well as disordered chemical potential $\mu_j \in [-\sqrt{3} \sigma_\mu, \sqrt{3}\sigma_\mu]$. For $\mu_j = 0$, zero modes exist for both $|U| > t$ and $|U| < t$, so the gap vanishes away from the transition. However, the accidental $|U| > t$ zero mode is destroyed by a non-zero chemical potential, whilst the topological mode $|U| < t$ persists. This is the case for uniform $\mu \not=0$ (a,b) and disordered $\sigma_\mu \not=0$ (c,d) chemical potential. 

By comparing to the exact results for the symmetric region, it is apparent that the topological phase boundary can be identified even for a small number of sites, and that the transition point depends smoothly and weakly on the chemical potential. This explicitly demonstrates that the ground-state degeneracy of the topological phase is stable for finite chemical potential, whilst that of the trivial phase is not. We therefore conclude that the qualitative aspects of the phase diagram (Fig.~\ref{figTopTrans}) are robust away from the symmetric region. In particular, as seen from (b) and (d), the phase transition occurs for an interaction strength greater than in the clean case $U = t$ for all $\mu$ and $\sigma_\mu$.

\emph{Discussion.---}
Our work provides a unique insight into the effects of disorder and interactions on topological order. The Kitaev chain in the symmetric region is equivalent to two copies of conventional non-interacting chains, see Eq.~\eqref{eqTransTF}. We have shown that topological order of the original system is related to that of the first of these copies (subsystem I), for which the interaction plays the role of the chemical potential. As a consequence, our analytic condition for the topological phase, Eq.~\eqref{eqMajNorm}, has a  similar form as that found for non-interacting, disordered Kitaev chains \cite{Gergs2016}. However, whilst the phase diagram of Ref.~\cite{Gergs2016} captures the competition between Anderson insulating and superconducting phases driven by chemical potential, we here study transitions driven by interaction. Whilst disorder and interactions can separately degrade the topological phase, their combination can be less detrimental.

Our explicit expression for the many-body Majorana mode \eqref{eqMajInteracting} is a rare example of an analytical expression of the general form in Eq.~(\ref{eqMajGeneral}), albeit with the simplification that most of the coefficients $\{\alpha\}$ are zero. This is due to the integrability of the system in the symmetric region. Nevertheless, one can see how higher-order multi-particle contributions occur at higher order in $U$ as expected from perturbation theory. In particular, we show explicitly that in the expansion of the many-body Majorana operator, terms with $(2n+1)$ $\gamma$-operators are proportional to $U^{n}$ \cite{Kells2015}.

Whilst our analytical results are restricted to the fine-tuned point, $t_j = \Delta_j$ and $\mu_j = 0$, our ED results demonstrate that the qualitative aspects of our findings hold more generally.

\emph{Conclusions and outlook.---}
We have been able to calculate analytically the topological phase boundary of a class of interacting, disordered Kitaev chains as a function of mean interaction and disorder strength, Fig.\@ \ref{figTopTrans}, demonstrating that moderate amounts of disorder in the interactions can enhance the topological phase into the strongly interacting regime $U>t$.

Our work represents a first step in utilizing the exact solution of Ref.\@ \cite{Miao2016} which enabled us to uncover an interacting part of the symmetric Kitaev chain phase diagram. Having understood the topological properties of the system, we can exploit it further to address a number of experimentally relevant questions. Specifically, how do interactions alter the zero-bias conductance peak \cite{Thomale2013} or topological Josephson current \cite{Fu2009}?

The analytic tractability of the symmetric Kitaev chain also holds great promise of studying the effects of interactions and disorder in a number of other situations. In particular, we suggest looking at the nature of localized states in the disordered chain in the context of many-body localization \cite{Nandkishore2015}. Additionally, the non-local nature of the transformation is likely to affect entanglement dynamics and out-of-equilibrium phenomena~\cite{Vasseur2014, Smith2017}. Finally, our exactly soluble interacting chain with disorder will provide a new benchmark point for numerical methods such as DMRG \cite{Gohlke2017}.

\emph{Acknowledgements.---}
J.~K.~is supported by the Marie Curie Programme under EC Grant Agreement No.~703697. A.~N.~holds a University Research Fellowship from the Royal Society and acknowledges support from the Winton Programme for the Physics of Sustainability.

\bibliography{paper}

\newpage
\appendix

\setcounter{figure}{0}
\makeatletter 
\renewcommand{\thefigure}{S\arabic{figure}}

\newcounter{defcounter}
\setcounter{defcounter}{0}

\newenvironment{myequation}{%
\addtocounter{equation}{-1}
\refstepcounter{defcounter}
\renewcommand\theequation{S\thedefcounter}
\begin{equation}}
{\end{equation}}

\begin{widetext}
\begin{center}
{\fontsize{12}{12}\selectfont
\textbf{Supplemental Material for ``Robustness of Majorana edge modes and topological order --- \\exact results for the symmetric interacting Kitaev chain with disorder''\\[5mm]}}
{\normalsize Max McGinley, Johannes Knolle, and Andreas Nunnenkamp\\[1mm]}
{\fontsize{9}{9}\selectfont  
\textit{Cavendish Laboratory, University of Cambridge, Cambridge CB3 0HE, United Kingdom}}
\end{center}
\normalsize
\end{widetext}

\section{Entanglement spectrum}

The ground-state density matrix for a system is given by the projector onto the lowest-energy eigenstate $\rho = \ket{\Psi_0}\bra{\Psi_0}$. Given a finite region $R$ of a large system, we can calculate the reduced density matrix $\rho_R$ of $R$ by taking $\rho$ and tracing out all the degrees of freedom not in $R$. This will yield an operator that characterizes the long-range entanglement between $R$ and its complement.
In the following we demonstrate how to calculate the eigenvalues of the entanglement Hamiltonian $\mathcal{H}_E = -\ln \rho_R$, which is known as the entanglement spectrum over $R$ \cite{Li2008}. Our method is based on that presented in Ref.~\cite{Latorre2004}, in which the entanglement spectrum of a spin-$\tfrac{1}{2}$ chain is calculated by means of a mapping onto free fermions. Their key observation is that for \emph{eigenstates} of a $Z_2^c$-symmetric Hamiltonian, tracing over a block of spins is equivalent to tracing over a block of fermions. In our case, the mapping in question is given by Eq.~\eqref{eqFullTrans} instead of the Jordan-Wigner transformation, however the symmetries of our Hamiltonian allow us to make a similar statement. The Hamiltonian commutes with both $Z_2^c$ and the particle-hole symmetry operator  \cite{Miao2016} 
\begin{align*}
Z_2^p \coloneqq \prod_{\substack{j=1 \\ j \, \text{odd}}}^{N-1} (i\gamma_j^A \gamma_{j+1}^B) = \prod_{\substack{j=1 \\ j \, \text{odd}}}^{N-1} (i\lambda_j^A \lambda_{j+1}^B).
\end{align*}
We can expand the reduced density matrix in the $\gamma$ basis as
\begin{align*}
\rho = 2^{-|R|}\sum_\mathcal{O} \braket{\Psi_0 | \mathcal{O} | \Psi_0} \mathcal{O}^\dagger
\end{align*}
where the operators $\mathcal{O}$ are a complete basis of operators in the subregion $R$ with the inner product $\Tr (\mathcal{O}_i^\dagger \mathcal{O}_j) = 2^{|R|}\delta_{i,j}$. Due to the symmetries of the Hamiltonian, if $\ket{\Psi_0}$ is an eigenstate then the only non-zero expectation values are those with an even number of $\gamma$ operators and an equal number of $\gamma^A_{2j + 1}$ and $\gamma^B_{2j}$ operators. One can show that such operators, when mapped into the $\lambda$ basis via \eqref{eqFullTrans}, have no strings extending outside the region $R$, and so tracing over $\gamma$ operators not in $R$ is equivalent to tracing over $\lambda$ operators not in $R$.

Additionally, as explained in the main text, see Eq.~(\ref{eqTransTF}), our model divides into two subsystems, and the condition for robust edge modes is that subsystem I (denoted with $\phi_j^{A,B}$ operators) is in the topologically non-trivial phase. Therefore we restrict the following calculation to the $\phi$ operators only.

Since the transformed Hamiltonian (\ref{eqTransTF}) is quadratic, due to Wick's theorem, the expectation values of any operator factorize into sums of products of two-fermion expectation values. Defining the correlation matrix $\Gamma$ of the system in the $\phi$-basis
\begin{align}
\braket{\Psi_0|\phi_j^\alpha \phi_k^\beta |\Psi_0} - \delta_{j,k}\delta^{\alpha, \beta} =  i \Gamma_{j,k}^{\alpha, \beta}
\label{eqCorrMat}
\end{align}
we can calculate any operator expectation value from products and sums of elements of $\Gamma_{j,k}^{\alpha, \beta}$. If we choose to trace over all degrees of freedom \textit{not} in the region $R$, then we only need to know $\Gamma_{j,k}^{\alpha, \beta}$ for $j$ and $k$ in $R$. The restriction of $\Gamma$ to $R$ is called the reduced correlation matrix $\Gamma^R$.

We can relate the reduced correlation matrix $\Gamma^R$ to the reduced density matrix $\rho_R$ by using $\braket{\Psi_0|\mathcal{O} |\Psi_0} = \Tr (\rho_R \mathcal{O})$ for any operator $\mathcal{O}$ depending only on degrees of freedom in $R$. As noted by Peschel \cite{Peschel2003}, the fact that all expectation values can be decomposed into Wick products means that $\mathcal{H}_E$ itself must be quadratic. Since $\mathcal{H}_E$ generates the correlation matrix \eqref{eqCorrMat}, $\Gamma$ and $\mathcal{H}_E$ must have the same eigenstates, so we have
\begin{align*}
\mathcal{H}_E &= \frac{i}{2} \sum_j \nu_j \chi_j^A \chi_j^B \\
\tilde{\Gamma}^R_{j,k} &\coloneqq -i\braket{\Psi_0|\chi_j^A \chi_k^B |\Psi_0} = \delta_{j,k} \eta_j 
\end{align*}
where $\{\chi_j^{A,B}\}$ are the Majorana operators in which $\mathcal{H}_E$ is diagonal, obtained by diagonalizing $\Gamma^R$, and where $\tilde{\Gamma}^R$ is the correlation matrix in this eigenbasis. Finally, by substituting $\rho_R = e^{-\mathcal{H}_E}$ into the above we obtain
\begin{align*}
\tanh \left(\frac{\nu_j}{2}\right) = \eta_j.
\end{align*}
If any of the $\nu_j$ are zero then the full entanglement spectrum is doubly degenerate, with equal-energy states related by `entanglement zero modes' $\chi_j^A$ and $\chi_j^B$.

\section{Disorder in the hopping term}

In this section we show that our results, which focussed on disorder in the interaction matrix elements, are robust against perturbations within the symmetric region.
We introduce additional disorder in the hopping amplitudes, in particular uniform disorder $t_j = \Delta_j \in 1 + [-\sqrt{3}\sigma_t, +\sqrt{3}\sigma_t]$.
As we are interested in perturbing the phase diagram shown in Fig.~\ref{figTopTrans}, the disorder strength in the hopping $\sigma_t$ and in the interaction $\sigma_U$ are chosen to be $\sigma_t = \alpha \sigma$ and $\sigma_U = (1-\alpha)\sigma$.

\begin{figure}
\includegraphics[scale=1]{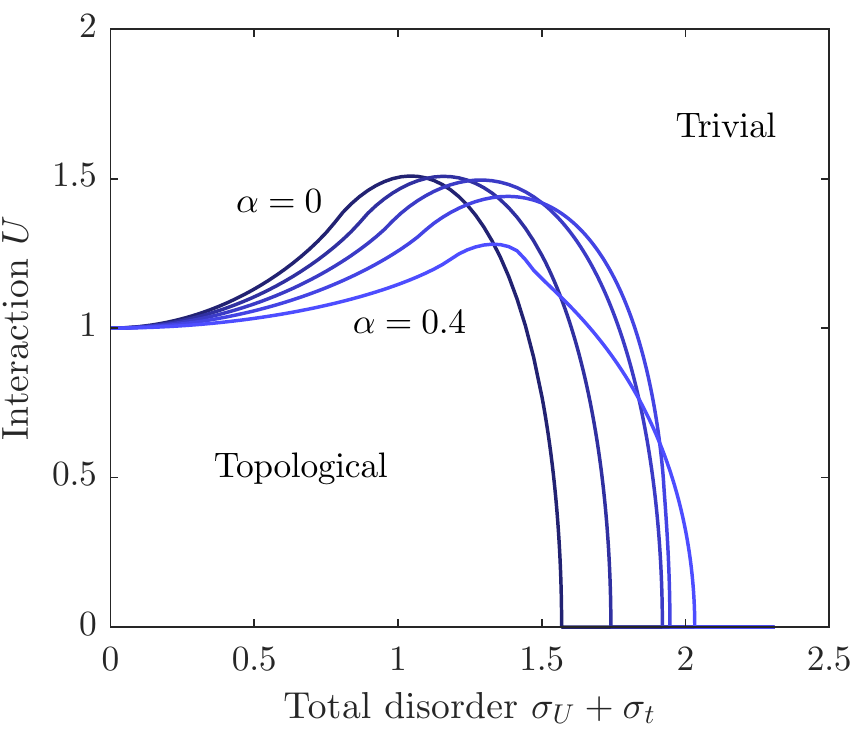}
\caption{Topological phase diagram of Hamiltonian \eqref{eqHamiltOrigGeneral} at the symmetric point as a function of the mean interaction strength $U$ and the total disorder strength $\sigma = \sigma_U + \sigma_t$. The phase boundary calculated from equation \eqref{eqMajNorm} is plotted for ratios of disorder strength $\alpha = \sigma_t / (\sigma_U + \sigma_t)$ increasing in steps of $0.1$ from $0$ to $0.4$.}
\label{figTDisorder}
\end{figure}

In Figure \ref{figTDisorder} we show the topological phase diagram in the infinite-system limit for various ratios of the disorder strength $\alpha$, as calculated from Eq.~\eqref{eqMajNorm}. As we gradually turn on hopping disorder by increasing $\alpha$, the maximal critical interaction strength moves smoothly to greater disorder. Additionally, the critical disorder strength increases smoothly. We see that even for significant hopping disorder ($\alpha = 0.4$) the qualitative features of the phase diagram in Fig.~\ref{figTopTrans} do not change -- the topological phase is enhanced by moderate disorder and destroyed by strong disorder.

\end{document}